\documentclass[twocolumn]{aastex61}

\received{18 November 2017}
\revised{-- March 2018}
\accepted{-- March 2018}
\submitjournal{The Astrophysical Journal}

\shorttitle{Excitation Temperatures of the OH 18-cm Main Lines in W5}
\shortauthors{Engelke and Allen}

\begin{document}

\title{OH as an Alternate Tracer for Molecular Gas: \\
Excitation Temperatures of the OH 18-cm Main Lines in W5}

\correspondingauthor{Philip Engelke}

\author[0000-0002-0786-7307]{Philip D. Engelke}
\altaffiliation{Grote Reber Fellow}
\affiliation{Department of Physics and Astronomy, The Johns Hopkins University, Baltimore, MD 21218, USA}
\affiliation{National Radio Astronomy Observatory, 1003 Lopezville Road, Socorro, NM 87801, USA}
\email{pengelk1@jhu.edu}

\author[0000-0001-9906-8352]{Ronald J. Allen}
\affiliation{Space Telescope Science Institute, 3700 San Martin Drive, Baltimore, MD 21218, USA}
\affiliation{Center for Astrophysical Sciences, Department of Physics and Astronomy, The Johns Hopkins University, Baltimore, MD 21218, USA}
\email{rjallen@stsci.edu}




\begin{abstract}

We present excitation temperatures $T_{ex}$ for the OH 18-cm main lines at 1665 and 1667 MHz measured directly in front of the W5 star-forming region, using observations from the Green Bank Telescope and the Very Large Array. We find unequivocally that $T_{ex}$ at 1665 MHz is greater than $T_{ex}$ at 1667 MHz. Our method exploits variations in the continuum emission from W5, and the fact that the continuum brightness temperatures $T_C$ in this nebula are close to the excitation temperatures of the OH lines in the foreground gas. The result is that an OH line can appear in emission in one location and in absorption in a neighboring location, and the value of $T_C$ where the profiles switch from emission to absorption indicates $T_{ex}$. Absolute measurements of $T_{ex}$ for the main lines were subject to greater uncertainty because of unknown effects of geometry of the OH features. We also employed the traditional ``expected profile" method for comparison with our ``continuum background" method, and found that the continuum background method provided more precise results, and was the one to definitively show the $T_{ex}$ difference. Our best estimate values are: $T^{65}_{ex} = 6.0 \pm 0.5$ K, $T^{67}_{ex} = 5.1 \pm 0.2$ K, and $T^{65}_{ex} - T^{67}_{ex} = 0.9 \pm 0.5$ K. The $T_{ex}$ values we have measured for the ISM in front of W5 are similar to those found in the quiescent ISM, indicating that proximity to massive star-forming regions does not generally result in widespread anomalous excitation of OH emission.

\end{abstract}




\section{Introduction}

\subsection{Background} \label{subsec:background}

This work is part of a broader project to study the suitability of the OH 18-cm transition as an alternate tracer for molecular gas in the Galaxy; see \citet[][]{arb12, arb13} for the initial motivation and \citet[][]{ahe15} for first results from this project. Tracing the molecular gas component of the ISM is complicated by the fact that the primary component of this gas, H$_2$, is not itself usually detectable in the conditions of interstellar molecular gas clouds, so signals from other molecules must be used as tracers. Despite its universal use as a tracer of cold molecular gas, the $^{12}$CO(1-0) transition is not an ideal tracer for molecular gas below the critical density of $\sim 10^3 $ cm$^{-3}/\tau$. Moreover, evidence from gamma-ray and IR surveys suggests the presence of undetected gas in the Galaxy, which is likely molecular and which likely contains a mass similar to the mass of known Galactic molecular gas \cite[]{gct05,t15}. In our work, OH serves as an alternate molecular gas tracer with a critical density of $\sim 10 $ cm$^{-3}$. We have found that for some molecular gas features, the OH and the CO traced the same amount of molecular gas and fell on a linear relation, whereas for other molecular gas features, the CO either under-predicted the molecular gas column density as compared to the OH results, or was not even detected above the noise. The conclusion in \cite{ahe15} was that OH traces a larger component of the molecular ISM than CO does, providing a means to probe the ``CO-dark'' molecular gas, and also has the benefit of being an optically thin line so column densities can be calculated directly.

One closely-related goal is to determine the molecular gas content in a star-forming region using OH emission, and to compare the values so obtained with those inferred using the CO(1-0) tracer for the molecular ISM. It is widely accepted that star-forming regions in galaxies contain an abundance of interstellar gas, and furthermore it is this higher gas abundance which is itself the cause of the elevated rate of star formation. The evidence for a close association of CO emission with star formation is primarily based on studies of active star-forming regions in our own Galaxy. However, we have described evidence that there is molecular gas in the ISM that is not traced by CO. Is there even more molecular gas hiding in star-forming regions, and could the 18-cm OH transition be used to find it? Or alternatively, is the molecular gas in star-forming regions adequately traced by CO emission because of the higher density of these regions, whereas it is only in quiescent regions in which a CO-dark component to the molecular gas exists? These questions motivate the application of the OH 18-cm transition as a molecular gas tracer to a star-forming region. To that end, a blind grid survey of the main OH lines at 1665 and 1667 MHz has been carried out over the star-forming region W5 with the Robert C. Byrd Green Bank Telescope (GBT). 

W5 was chosen as a target for our grid survey because it is a star-forming region of modest angular size compared to the resolution of the GBT, allowing lengthy 2-hour observations in a dense grid of $\sim 75$ pointings over the nebula to be carried out in a feasible amount of observing time. In addition, extensive results on this nebula are available in the literature including radio continuum data provided by the Canadian Galactic Plane Survey \cite[]{tgp03} and used in a detailed study of the properties of the W3-W4-W5 complex \cite[]{norm97}. That study shows that the brightness of the radio continuum emission emanating from W5 has values small enough to allow OH emission as well as absorption to be detected in the foreground gas. Finally, W5 is sufficiently simple in structure and activity so as to be tractable for analysis, and does not contain large numbers of masers or copious IR radiation which might adversely affect our ability to analyze the OH L-band emission in terms of column density in the ISM.


However, before column densities of OH can be computed from those observations, accurate values for the excitation temperatures $T_{ex}$ of the OH 18-cm lines in W5 must be known. Knowledge of $T_{ex}$ is particularly important when the OH emission is found in regions with star formation activity, because the excitation temperature of the L-band lines is known from previous work to be only a few degrees above, or comparable to, the background value established by the sum of the Cosmic Microwave Background (CMB), the Galactic non-thermal radio continuum emission, and the thermal radio continuum emission from the star-forming region itself. The difference between the total ambient L-band radiation field and the excitation temperature is one factor that determines the intensity of the observed emission lines, and hence the column density of OH obtained from the observations. This paper focuses on measurements of $T_{ex}$ for the two main OH lines in order to permit accurate determinations of OH column densities in the ISM. The column density results will be reported in another paper that is in preparation.

\subsection{OH Excitation Temperatures} \label{subsec:Tex}

The two main 18-cm lines of OH at 1665 MHz and 1667 MHz were first detected in absorption in the ISM by \cite{wbmh63} in front of the continuum source Cas A. OH in emission was first reported by \cite{wwdl65} in an HII region, and then by \cite{h68} in interstellar dust clouds. An extensive survey of northern hemisphere radio sources for OH absorption in the Galaxy was carried out by \cite{g68}. At first, it was generally assumed that $T_{ex}$ was the same for the two main lines at 1665 MHz and 1667 MHz; \cite{h69} reported that his observations did not show evidence for a difference between the main-line excitation temperatures.

During the next few years, it was unclear if $T_{ex}$ was the same at the two main lines, or if there were differences. Some evidence for $T_{ex}$ differences was found by \cite{mg71}, and \cite{t73}, although those results were disputed \cite[e.g.][]{hg75}. Moreover, these results varied from finding evidence that $T_{ex}$ was larger at 1667 MHz to finding that $T_{ex}$ was larger at 1665 MHz. \cite{rwg76} found evidence that $T^{65}_{ex} - T^{67}_{ex}$ of approximately 1.5 K, but with sufficient uncertainty as to claim that the difference was ``not significant.''  \cite{c77,c79} found more definitively that $T^{65}_{ex} > T^{67}_{ex}$ in dust clouds in front of several continuum sources at a number of LSR radial velocities between -10 km/s and 10 km/s, all closer to the sun than in the present study of W5. \cite{c79} reported that $T^{65}_{ex} - T^{67}_{ex} = 1-2$ K, and also described an observation in which OH emission was detected at 1665 MHz while absorption was detected at the same position at 1667 MHz, providing a $4\sigma$ qualitative result that $T^{65}_{ex} > T^{67}_{ex}$ at that location in the sky near 3C 154 and at -1 km/s LSR radial velocity. Such a difference was not, however, reported by \cite{ll96}. \cite{li18} provide a distribution of $T_{ex}$ measurements, with a peak near 3.4 K but an average value of the measurements ranging from approximately 4.8 K to 7.0 K, their exact average value depending on whether the measurements are weighted by the errors when averaging, and with an excitation temperature main line difference no greater than 2 K. A table of $T_{ex}$ measurements from the literature is assembled in the Appendix.

\subsection{Another Approach} \label{subsec:NewApproach}

The measurement of excitation temperatures has traditionally employed what we shall call the ``expected profile'' method, where an estimate of the expected emission profile at the sky position of the distant source of continuum emission is provided by interpolating from the emission profiles recorded at offset positions around the location of the continuum source. This method is subject to several uncertainties caused by spatial gradients in the foreground absorbing gas and differences in the effective solid angle covered by the distant radio continuum source and the telescope used to determine the expected profile. This method has nevertheless been used by many researchers dating back to the early observations \cite[e.g.\,][]{mg71, c72, rwg76}; see \citet[][]{ll96} for a more detailed description of the method.
 
In the course of analyzing the GBT W5 data, we have implemented an alternative method for measuring the excitation temperatures of the OH lines. We make use of the natural variations in the 18-cm radio continuum brightness over W5 and the fact that the values of the continuum brightness temperature $T_C$ in this nebula are close to the excitation temperature of the main OH lines in the foreground gas. The key is that if $T_C$ is less than $T_{ex}$, lines will appear in emission, if $T_C$ is greater than $T_{ex}$, lines will appear in absorption, and if $T_C$ = $T_{ex}$, the lines will disappear. Some aspects of this approach were also used by \cite{nkg02}, and the reasoning that $T_C$ where an emission line disappears and then switches to absorption indicates $T_{ex}$ was discussed by \cite{t17}. This method, which we call the ``continuum background'' method, unequivocally shows that the OH molecular gas in front of the W5 continuum source has $T^{65}_{ex} > T^{67}_{ex}$. 

In order to compare these results with the traditional expected profile method, we have supplemented the GBT observing program with OH absorption measurements on extragalactic radio sources seen through the radio image of W5; these absorption observations have been done with the Karl G. Jansky Very Large Array (VLA). In the following section we describe both of these interconnected observing programs.

\section{Observations} \label{sec:observations}

\begin{figure*}[ht!] 
\vspace{0.1in}
\begin{center}
\includegraphics[width=1.05\textwidth, angle=0]{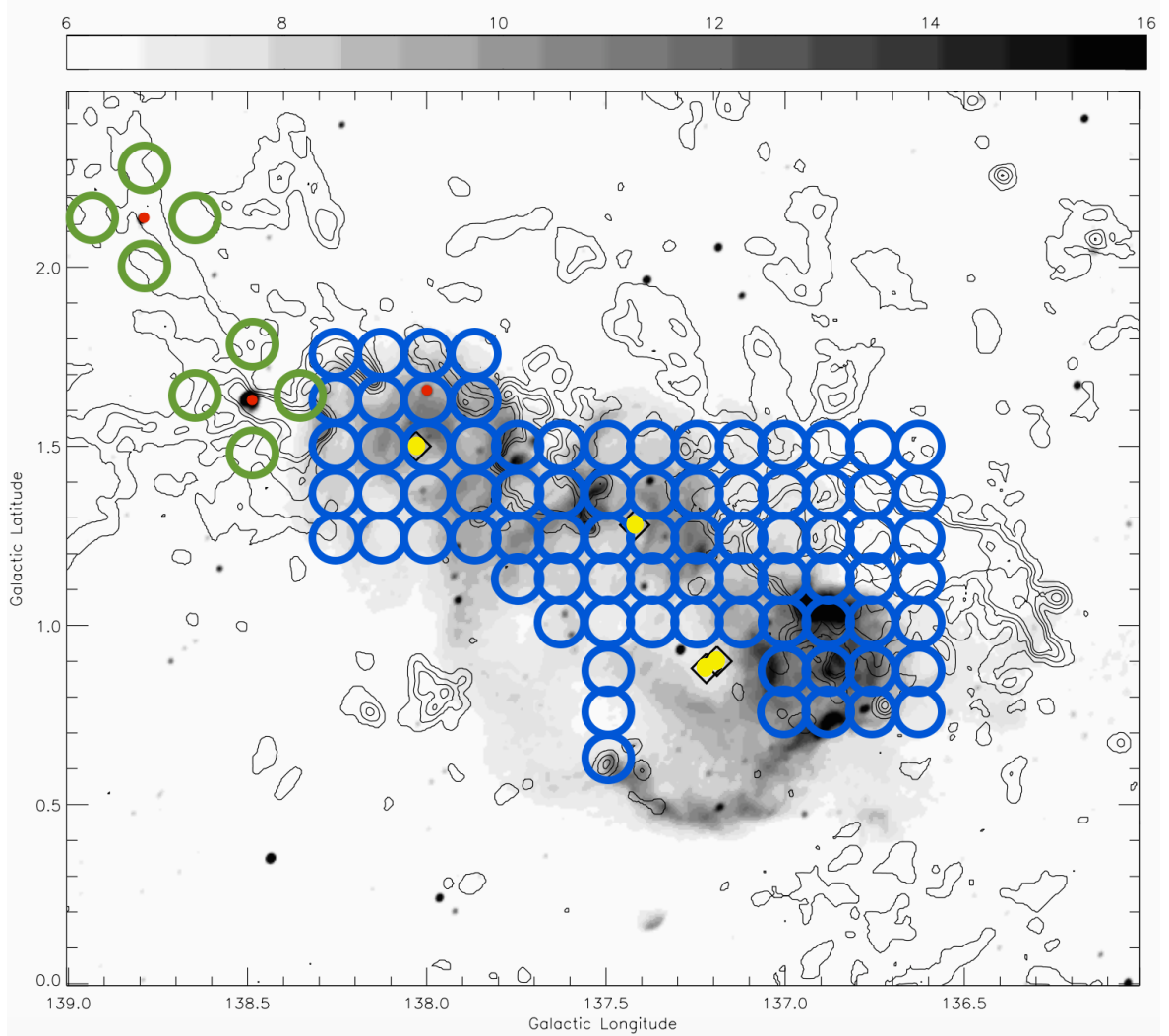} \hspace{0.2in} 
\caption{Positions of our GBT and VLA observations overlaid on the map of the W5 region taken directly from \citet[][Figure 2]{km03}. The gray scale indicates the 1420 MHz continuum, the contours indicate CO emission in the LSR radial velocity range of -49 to -31 km s$^{-1}$, and the 4 diamonds (which we have emphasized in yellow) indicate the locations of known O stars. On top of the map we have plotted a number of blue circles indicating the positions of our GBT observations in the grid survey, with the circle size equivalent to the point spread function FWHM of $7\farcm 6$. The 3 red dots indicate the positions of our VLA observations, and the two sets of 4 green circles indicate the positions of our GBT observations in the vicinity of two of the 3 VLA observations for expected emission profile detection (the third set of 4 pointings is already covered in the grid of blue circles). \label{fig:W5-region}}
\end{center}
\end{figure*}

Observations for this work made use of the 100-m Green Bank Telescope (GBT) in West Virginia and the Very Large Array (VLA) in New Mexico. The GBT was used for both methods of determining the OH excitation temperatures; the VLA data was used only for the expected profile method. The grid of GBT pointings and the locations of the 3 NVSS sources observed with the VLA are shown in Figure \ref{fig:W5-region} superposed on a drawing from  \citet[][their Figure 2]{km03} of the distributions of radio continuum (gray scale) and CO(1-0) emission (contours) in the region of W5.

All GBT observations were performed with the VEGAS spectrometer in L-band, with 2 hours of exposure per pointing in frequency-switching mode. All four OH 18-cm transitions at 1612 MHz, 1665 MHz, 1667 MHz, and 1720 MHz, plus 1420 MHz HI were recorded simultaneously; however, only the 1665 MHz and 1667 MHz ``main-line'' spectra were used in this work. The beam size of the GBT at 1667 MHz was $7\farcm 6$ FWHM, and the spectrometer channel spacing was 0.515 kHz, corresponding to 0.0926 km s$^{-1}$ at 1667 MHz. The GBT spectra were smoothed off-line to a Gaussian FWHM resolution of 1.0 km s$^{-1}$. The RMS noise level in the final smoothed spectra was 0.0028 K per channel. A total of 80 GBT pointings was made in the W5 grid survey between February 2016 and April 2017, and an additional eight GBT observations were made between February 2017 and August 2017 in order to construct the expected profiles near 2 of the 3 compact continuum sources from the NVSS catalog \citep[][]{ccg98} observed with the VLA in the immediate area of W5. The third NVSS source turned out not to be suitable for the expected profile method because its surroundings contained $T_C > T_{ex}$. 

VLA observations of three compact continuum sources NVSS J030317-602752 (Source 1), NVSS J030723-604522 (Source 2), and NVSS J025947-604325 (Source 3), in or near the radio image of W5 were performed in the D configuration as well as the DnC hybrid configuration with the L-band receiver, providing a resolution of $62\arcsec$. The channel separation in the spectrometer was 10.845 kHz corresponding to 1.95 km s$^{-1}$ at 1667 MHz. These three sources were observed in April 2017 and May 2017, with different exposure times depending on the flux density of the source. Unfortunately, owing to limitations on the observing time available, only the brightest source received the full intended exposure times of 3.3 hours; Source 2 received 17.6 hours, and Source 3 received 18.7 hours.

\section{Results and Analysis} \label{sec:analysis}

Both of the methods we use to measure excitation temperatures begin with the simplest form for the equation of radiative transfer for a uniform slab:
\begin{equation}
T_L = (T_{ex} - T_C)(1 - e^{-\tau}), \label{eqn:radx}
\end{equation}
where $T_C$ is the spectrum of the total continuum emission incident on the back of the slab, $T_{ex}$ is the excitation temperature of gas in the slab for the  level under study, $T_L$ is the observed emission profile exiting the slab, and $\tau$ is the absorption line optical depth through the slab. This equation immediately suggests the solution for the phenomenon described in Section 1.3; if $T_{ex} > T_C$ the spectral line exiting the slab will be seen in emission, but if $T_{ex} < T_C$ it will be seen in absorption. Crucially, if $T_{ex} = T_C$, the emerging profile will disappear. If the value of $T_C = T_{CMB} + T_{Gal} + T_{W5}$ at that point is known to some precision, the value of $T_{ex}$ can be determined with the same precision. Furthermore, the difference in the values of $T_C$ where the two main lines disappear is the difference in their excitation temperatures. Since the values of $T_{CMB}$ and $T_{Gal}$ are expected to vary only slowly over the radio image of W5, the difference in the excitation temperatures of the two main lines $T^{65}_{ex} - T^{67}_{ex}$ depends on the variations in $T_{W5}$, removing the necessity for an absolute measurement of continuum temperatures, but requiring an accurate relative continuum surface brightness distribution. This continuum image will be described next.

\subsection{A Continuum Image of W5 at 1667 MHz} \label{subsec:W5Continuum}

In order to maximize sensitivity, the GBT observations were made using frequency-switching instead of position-switching. Unfortunately, such observations do not provide information about the continuum temperatures. We have interpolated observations from the Canadian Galactic Plane Survey at 408 MHz and 1420 MHz \cite[]{tgp03} and the Effelsberg survey at 2695 MHz \cite[]{frrr90} from their observed frequencies to the main OH lines assuming a multi-component radio spectrum including background contributions and contributions from W5, and we smoothed the result to the spatial resolution of the GBT observations. However, it was first necessary to separate the free-free emission of W5 from the cosmic microwave background and the Galactic background. The Effelsberg data defined the background in the W5 vicinity as 0 K, so only the W5 free-free emission was included in that data. On the other hand, the Canadian Galactic Plane Survey data gives the absolute continuum temperatures, meaning that the background in the vicinity of W5 is not zero, and includes contributions from the CMB and the Galactic background. Our estimates for the background at 408 MHz and 1420 MHz contribute some uncertainty to the results for $T_C$, estimated as $\pm 7$K at 408 MHz and $\pm 0.25$K at 1420 MHz. A linear fit was made to the logarithm of the $T_C$ values plotted against the logarithm of the frequencies, yielding a spectral index of -$1.8\pm 0.15$, which is consistent with theoretical expectations for free-free emission. We then interpolated from 1420 MHz $T_C$ data smoothed to the resolution of the GBT observations using the spectral index of -1.8 to find $T_C$ at 1667 MHz. The uncertainty in $T_C$ at 1667 MHz is $\pm$ 0.07 to 0.1 K, depending on the magnitude of $T_C$. 

\subsection{Difference in Excitation Temperature Between the Main Lines}

\begin{figure*}[ht!] 
\vspace{0.1in}
\begin{center}
\includegraphics[width=0.7\textwidth, angle=0]{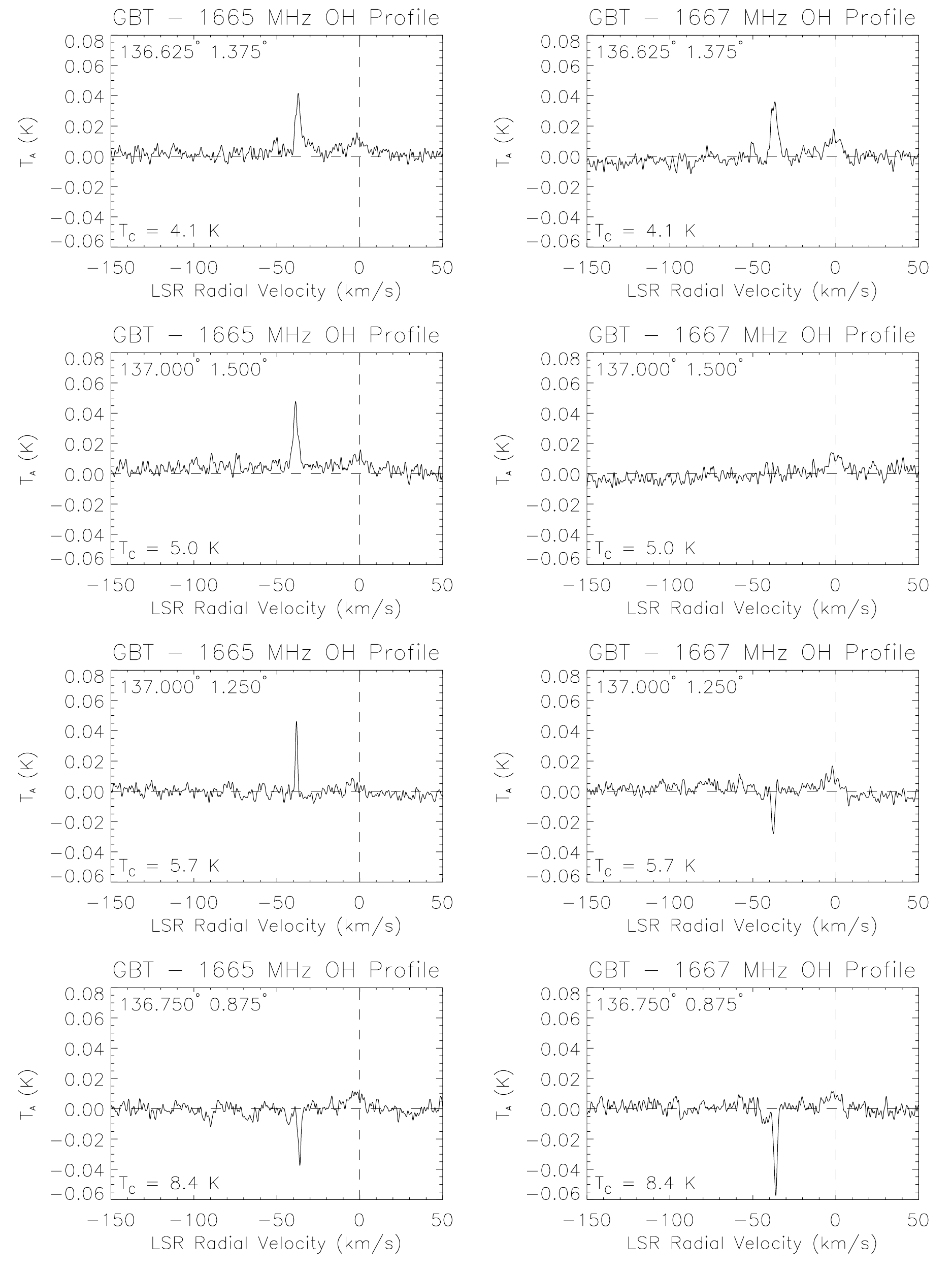} \hspace{0.2in} 
\caption{Illustration of excitation temperature estimation using the continuum background method with OH spectra from the GBT survey of W5. The spectra at each row are from different positions in W5 with a continuous increase in continuum temperature going downward in the figure. The spectra on the left are at 1665 MHz, and the spectra on the right are at 1667 MHz. As the continuum temperature gradually increases, OH features eventually change from emission to absorption. The continuum temperature at which the switch from emission to absorption occurs indicates the excitation temperature of the OH line. The excitation temperature is not the same at 1665 MHz and at 1667 MHz. This point is made clearly by the fact that whereas the 1667 MHz spectra change from emission to no apparent signal to absorption from the first to the third rows downward, the 1665 MHz spectra at these same positions (and with the same continuum temperatures) are all emission lines. Not until the bottom row does the 1665 MHz line change to absorption, which occurs at a higher continuum temperature. \label{fig:SmokingGun}}
\end{center}
\end{figure*} 

One benefit of using the continuum background method is that this method demonstrates clearly that $T^{65}_{ex} >T^{67}_{ex}$. At five positions with the appropriate continuum temperature, the 1665 MHz line appears in emission while the 1667 MHz line at the same position, with the same continuum background and OH content, appears in absorption (see Figure \ref{fig:SmokingGun} for an example). The observations make sense in the context of a difference in $T_{ex}$ values between the two main lines. In that case, for the same $T_C$, $T^{65}_{ex} >T_C$ so the line appears in emission, and $T^{67}_{ex} < T_C$, so the line appears in absorption. Then, for a higher value of $T_C$, we find cases in which the 1665 MHz line and 1667 MHz line are both absorption lines, indicating that $T_C$ in these cases is greater than $T^{65}_{ex}$. Similarly, we find cases in which the 1665 MHz line and 1667 MHz line are both emission lines, indicating that $T_C$ in these cases is less than $T^{67}_{ex}$. 

Considering that the qualitative result that $T^{65}_{ex} > T^{67}_{ex}$ depends only on the emission and absorption nature of the line profiles, we can say that the confidence of this result is equivalent to the signal-to-noise of the line detections in the GBT observations. That means that in contrast to previous findings that $T^{65}_{ex} > T^{67}_{ex}$ in the literature, which depended on separate measurements of the values of $T_{ex}$ for the two main lines and therefore was subject to the uncertainty in those measurements, we report a confidence that $T^{65}_{ex} > T^{67}_{ex}$ in the W5 region to $7 \sigma$, and we have observed five examples of this type of spectrum (see Figure \ref{fig:ColorCode}). 

\subsection{Analysis using the Continuum Background Method}

Whereas the major source of uncertainty for the expected profile method is unknown variation in OH distribution in the tangential direction, the major source of uncertainty for the continuum background method is the unknown distribution of OH emission along the line of sight. A simplifying assumption that can be made for the continuum background method is that OH must be located entirely in front of or behind the continuum emission in W5 because otherwise OH would be dissociated in the HII region by high energy photons. This might be a reasonable assumption, but it is based on a simplified model of the geometry. In reality, HII regions and molecular clouds are not spherical, and fractal-like geometry paired with shielding effects could complicate this model significantly. 

\begin{figure*}[ht!] 
\vspace{0.1in}
\begin{center}
\includegraphics[width=1.0\textwidth, angle=0]{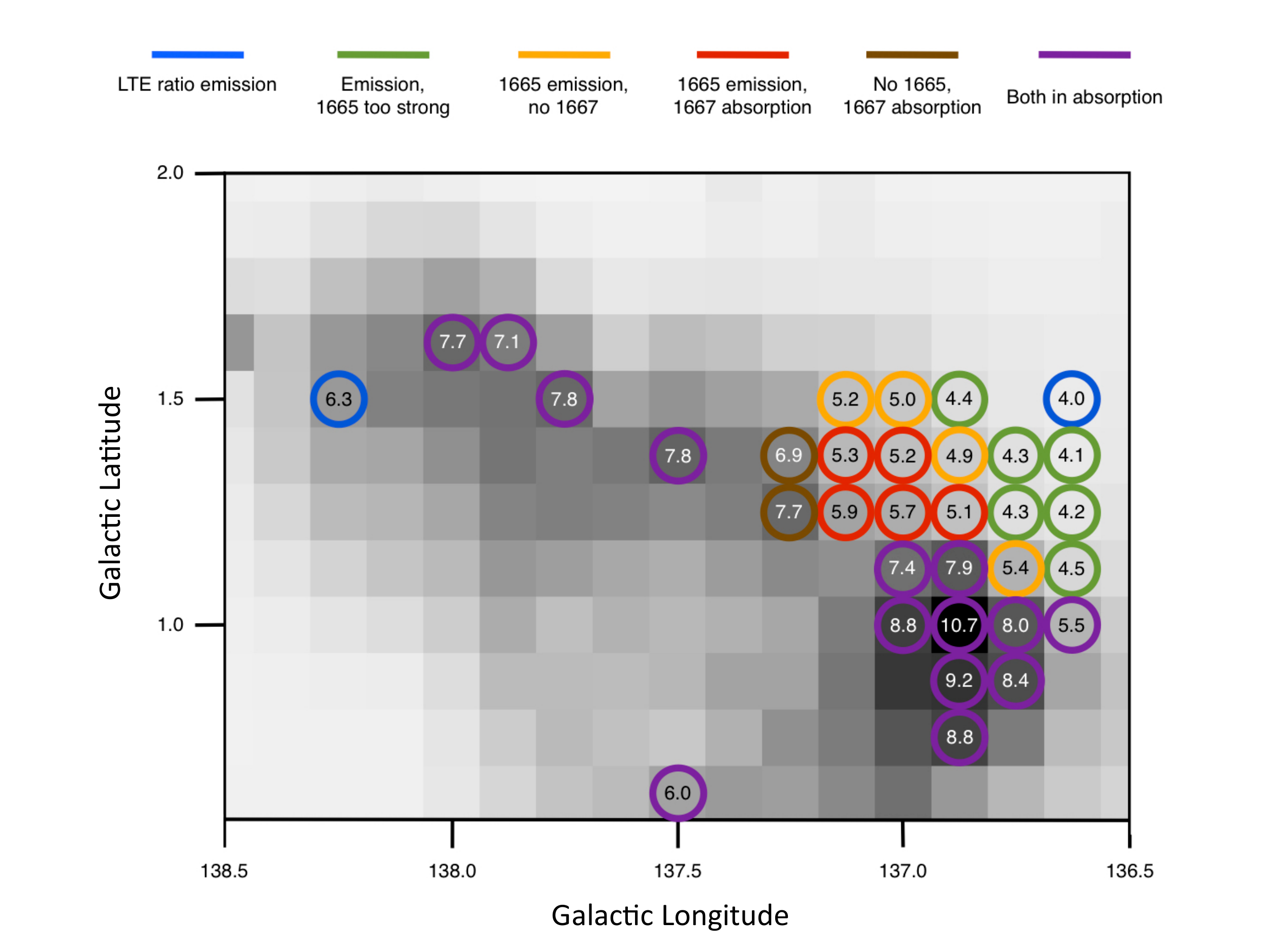} \hspace{0.2in} 
\caption{Map of 1420 MHz continuum in W5 smoothed to GBT survey resolution from the Canadian Galactic Plane Survey \cite[]{tgp03} in gray, with all positions observed in the GBT OH survey that contain OH feature detections at W5 radial velocities indicated with colored circles.  The colors refer to the types of OH features observed at 1665 and 1667 MHz, as indicated in the key, and the numbers written inside of the circles are the values of the continuum temperature at 1667 MHz for the GBT survey resolution.  Excitation temperatures for the 1667 MHz and 1665 MHz lines can be estimated from the continuum temperatures at the orange-circled positions, and the boundary between red- and purple-circled positions, respectively.}
\label{fig:ColorCode}
\end{center}
\end{figure*}

In order to determine values of $T^{65}_{ex}$ and $T^{67}_{ex}$, we produce a map of the survey region and color code all coordinates with OH detections at W5 radial velocities based on the types of OH detections that were observed at those coordinates, as seen in Figure \ref{fig:ColorCode}. We start by accepting the assumption that OH is located entirely in front of or behind the free-free continuum source along the line of sight. Looking at Figure \ref{fig:ColorCode}, we identify four estimates of $T_{ex}$ at 1667 MHz from the orange-circled coordinates, and taking the mean with standard deviation as the uncertainty, we obtain $T^{67}_{ex} = 5.1 \pm 0.2$ K. Uncertainty resulting from possible variation in $T_{ex}$ along the lines of sight is unlikely to be greater, and if it were, a combination of emission and absorption in a single profile would probably be visually apparent. The uncertainty in $T_C$ does not increase the total uncertainty by more than a few tenths of a degree when summed in quadrature. Determining $T^{65}_{ex}$ is a bit more difficult because there are only two coordinates with observations of a non-detection at 1665 MHz and an absorption detection at 1667 MHz, and this OH is evidently located farther back along the line of sight than most of the other OH detections. We surmise that it is located farther back because $T_C$ at these two brown-circled pointings varies by 0.8 K, and also both of these pointings have higher apparent $T_C$ than some of the purple-circled pointings. We already know that the assumption that OH is located in front of or behind the continuum source is not always true, so the following estimates are actually upper limits. Looking at the boundary between red- and purple-circled coordinates in Figure
\ref{fig:ColorCode}, we estimate a value of $T_{ex} = 6.0 \pm 0.5$ K, because it seems evident from the red-circled observation at 5.9 K and the purple-circled observation at 6.0 K ($l = 137.5^{\circ}$, $b = 0.625^{\circ}$) as well as various red-circled observations at lower $T_C$ values and purple-circled observations at higher $T_C$ values, that the boundary between red- and purple-circled positions occurs where $T_{C}$ is around 6.0 K. On the other hand, there is a stray purple-circled observation at 5.5 K, so we choose an uncertainty value of $\pm 0.5$ to cover this variation. The variation is most likely the result of a more pronounced absorption profile at the left edge of the GBT beam dominating over a weaker emission profile in the remainder of the beam, although the possibility that it is evidence of $T_{ex}$ variation, uncertainty in $T_C$, or the presence of OH in shielded pockets within the HII region at other observed positions should be noted. Our estimates using this method are:

\begin{center}
\begin{tabular}{c}
$T^{65}_{ex} = 6.0 \pm 0.5$ K, \\
$T^{67}_{ex} = 5.1 \pm 0.2$ K, \\
$T^{65}_{ex} - T^{67}_{ex} = 0.9 \pm 0.5$ K. 
\end{tabular}
\end{center}

Note that the blue-circled observation at $l = 138.25^{\circ}$, $b = 1.5^{\circ}$ can be understood by remembering that OH features may be found in front of or behind the HII region associated with W5. This case is probably simply one where the OH is behind the HII region, so the $T_C$ value behind the OH is actually about 4.1 K rather than 6.3 K.

More complicated line of sight geometry could result in greater uncertainties than reported in the estimates. If we cease to assume that all OH must be either in front of or behind the continuum emission (as we already know is not the case at the two brown-circled positions), then $T_C$ behind the observations could be lower than we assumed. In that case, the estimates reported above are upper limits on $T_{ex}$. The lower limit would be the $T_C$ background from the CMB and the Galaxy, which amounts to $\sim 4$ K in this region, giving us:

\begin{center}
\begin{tabular}{c}
4 K $ < T^{65}_{ex} < 6.5 $ K, \\
4 K $ < T^{67}_{ex} < 5.3 $ K, \\
$T^{65}_{ex} - T^{67}_{ex} \le 0.9 \pm 0.5 $ K. 
\end{tabular}
\end{center}

\section{The Expected Profile Method} \label{sec:ExpectedProfile}

The ``continuum background'' method described previously for measuring $T_{ex}$ has an attractive simplicity for the specific case of OH measurements. However, it is clearly not applicable in general; suitable extended continuum sources are not available everywhere on the sky, nor do they always have continuum surface brightnesses similar to the excitation temperature of the ISM tracer in question. In any case, it is also important to establish that the results of the new method are in agreement with those obtained with the more common ``expected profile'' method. This section explores that comparison.

The ``expected profile'' method of measuring excitation temperatures proceeds as follows: if absorption and emission lines are both detected in close proximity, the emission line temperature profile $T_L$ (i.e. the ``expected profile'') and the absorption line optical depth $\tau$ can be used to estimate $T_{ex}$ of the OH lines in the region. Assuming small optical depth we can write Equation \ref{eqn:radx} as:

\begin{equation}
T_{ex} \approx \frac{T_L}{\tau} + T_C.
\label{eqn:Tex-expected}
\end{equation}

\noindent where $T_L$ is the expected profile of the emission line at the location of the continuum absorber. This profile is obtained by averaging the emission profiles observed in regions immediately adjacent to the background radio continuum source, $T_C$ is the continuum temperature in the observations of the emission expected profile, and $\tau$ is the absorption line optical depth in front of the background continuum source.

\subsection{Analysis using the Expected Profile Method} \label{sec:ExpProfAnalysis}

We begin by noting the assumptions involved in using the expected profile method. We assume that $T_{ex}$ for a given line is constant throughout the region. This may not in fact be true, but we have found no evidence for significant differences. A second assumption is that the OH column density does not vary tangentially over the region around each continuum (VLA) source where the GBT expected profile observations were performed. In reality, this assumption is almost certainly violated, as can be inferred e.g.\ from the rapid variations in CO column densities in those regions shown in Figure \ref{fig:W5-region}, and is likely to dominate the uncertainties. We will estimate errors resulting from this variation as we perform the analysis. Given this complication, it is remarkable that \cite{c79} was able to detect that $T^{65}_{ex} - T^{67}_{ex} = 1-2$ K using the expected profile method; one possible reason is that he was observing OH features which are located closer to the sun, probably within $\sim $ 0.5 kpc according to their radial velocities, and hence cover solid angles on the sky ten to twenty times that of ours. The \cite{c79} beam size of 22' diameter covered an area roughly comparable to the size of the region covered by our four GBT observations for the expected profile of each NVSS source, which has a diameter of 25.6', so it makes sense that there would be less variation in OH column density within the region observed by \cite{c79} than in ours.

\begin{figure*}[ht!] 
\vspace{0.1in}
\begin{center}
\includegraphics[width=0.7\textwidth, angle=0]{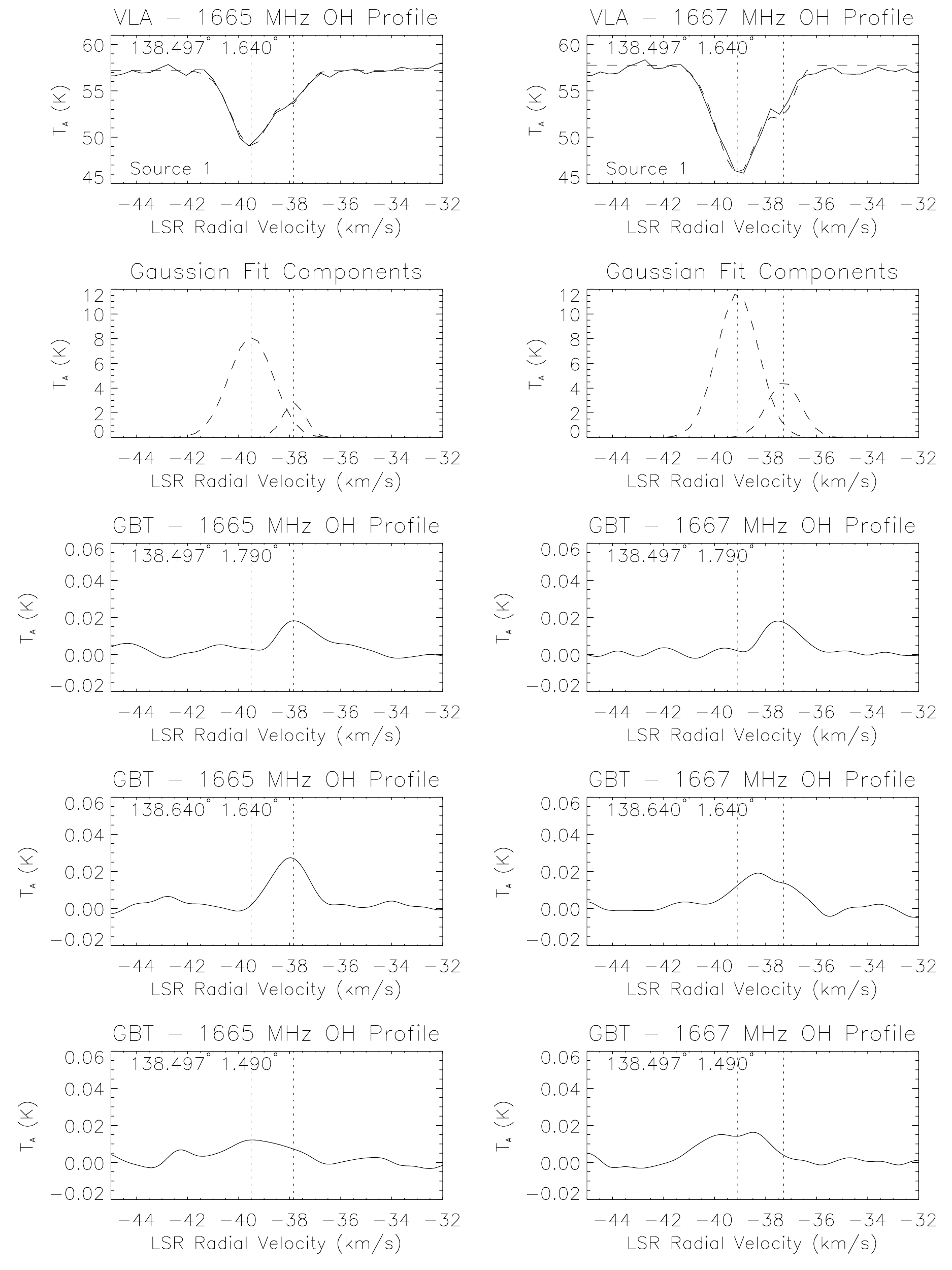} \hspace{0.2in} 
\caption{VLA absorption spectra from Source 1 and GBT emission spectra from its vicinity for the expected profile. The VLA absorption spectra are fitted with a two-component Gaussian, which is superimposed on each VLA spectrum as a dashed line. The two components of the fit are shown separately in the second row, as a dashed line, and the velocities at the centroids of these Gaussian fit components are marked on all rows as vertical dotted lines. Variation in GBT spectra among rows two, three, and four demonstrates some of the variations in OH content and distribution over the region that increases the uncertainty in $T_{ex}$ determined with the expected profile method. Also worth noting is the unusual case in the fourth row in which the 1665 MHz line appears to be brighter than the 1667 MHz line and the secondary component appears to dominate. This deviation is within the uncertainty so it could be attributed to noise, or perhaps it is evidence of a localized anomaly in the secondary component. \label{fig:Source1}
}
\end{center}
\end{figure*}

\begin{figure*}[ht!]
\vspace{0.1in}
\begin{center}
\includegraphics[width=0.8\textwidth, angle=0]{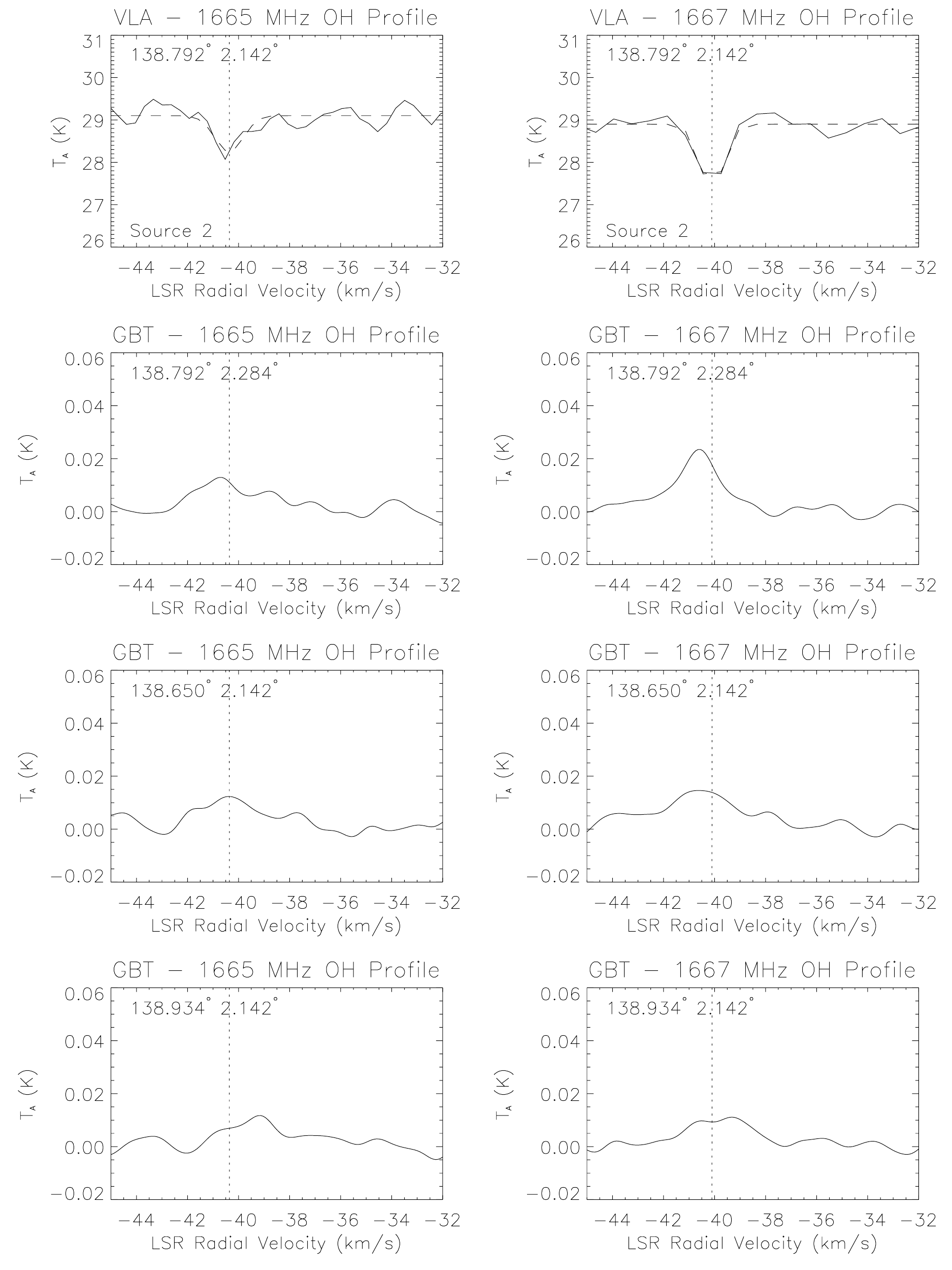} \hspace{0.2in} 
\caption{VLA absorption spectra from Source 2 and GBT emission spectra from its vicinity for the expected profile. Gaussian fits are performed on the VLA spectra and shown superimposed with a dashed line, and the centroid velocity of the Gaussian fit is displayed as a vertical dotted line on each spectrum. \label{fig:Source2}}
\end{center}
\end{figure*}

Of the three NVSS extragalactic continuum sources observed with the VLA over the area of W5, only two were useful for the analysis because $T_C$ turned out to be greater than $T_{ex}$ in the vicinity of the third source. The two remaining sources are located at $l = 138.497^{\circ}, b = 1.64^{\circ}$ (Source 1), and at $l = 138.792^{\circ}, b = 2.142^{\circ}$ (Source 2). The absorption spectra of these sources observed with the VLA at 1665 and 1667 MHZ are shown in the top pair of panels in Figure \ref{fig:Source1} and Figure \ref{fig:Source2}.

VLA observations of Source 1 as well as three of the four nearby GBT observations for the expected emission profile are shown in Figure \ref{fig:Source1}. The fourth GBT observation was not used because $T_C$ at that position was too high. The VLA observations at Source 1 reveal the presence of two OH components: a primary component centered near -39 km/s (called Component A), and a secondary component centered near -37.5 km/s (called component B). The expected profiles made by averaging the GBT spectra from observations surrounding Sources 1 and 2 respectively also display these two components, but in the case of the individual GBT spectra averaged, both components are not always detected together. As a result, it makes sense to treat the two components as separate OH features, fit the VLA absorption data as a sum of two Gaussian components, and compare the optical depth and line temperature values for the corresponding components when using Equation \ref{eqn:Tex-expected} to estimate the excitation temperatures. These $T_{ex}$ values can then be averaged together after their separate  determinations, but if the components of the OH features are not treated separately, the analysis will be faulty. The results of this exercise are given in Table \ref{table:source1compsA}. The uncertainties reported in this Table are calculated only from the error in the profile integral and the absorption optical depth measurements.

In order to estimate the error resulting from variation in OH, we look at some of the variation in OH profile integrals between adjacent pointings within our survey. We choose an example with background $T_C = 4.0$ K in order to best approximate the regions surrounding Sources 1 and 2, which have similar background $T_C$ values, because we wish to avoid making assumptions about $T_{ex}$ and how that would affect emission profile integral strengths. Between coordinates $l = 138.625^{\circ}, b = 1.5^{\circ}$ and $l = 138.75^{\circ}, b = 1.5^{\circ}$, the profile integral at W5 radial velocities varies between a nondetection and a detection, thus providing what seems like a decent estimate of one of the more rapid OH profile integral variations at approximately the relevant angular scale and continuum temperature. Using this estimate of the spatial variation of OH, we find an uncertainty of roughly $\pm 1.5$ - 2 K.

\begin{table*}[ht!]
\begin{center}
\caption{Excitation temperatures for the components of Source 1. Errors from profile integrals and optical depth measurements. \label{table:source1compsA}}
\begin{tabular}{c c c c c c}
\hline
 l & b & Profile Component & $T_C$ (K) & $T^{65}_{ex}$ (K) & $T^{67}_{ex}$ (K)\\
\hline
138.497$^{\circ}$ & 1.79$^{\circ}$ & B & 4.1 & 4.9 $\pm$ 0.3  & 4.5 $\pm$ 0.2 \\
138.647$^{\circ}$ & 1.64$^{\circ}$ & A \& B & 4.1 & 4.3 $\pm$ 0.2 & 4.2 $\pm$ 0.2 \\
138.497$^{\circ}$ & 1.49$^{\circ}$  &  A  &  4.2 &  4.4 $\pm$ 0.2 & 4.3 $\pm$ 0.2 \\                                    
\end{tabular}
\end{center}
\end{table*}
%
%


Taking the weighted mean of the results from the three positions (using the initial uncertainties before considering possible OH content variation over the region for the weighting), we have for the ``expected profile method'' result for Source 1 as follows: \\

4.1 K $ < T^{65}_{ex} <$ 6.5 K, and  4.1 K $< T^{67}_{ex} <$ 6.3 K \\

\noindent where $T_C$ = 4.1 K is a lower limit for both $T^{65}_{ex}$ and $T^{67}_{ex}$. \\

The VLA observations of Source 2 do not contain multiple detected OH components. The VLA absorption data were therefore treated as a single Gaussian component centered near -40 km/s. Three out of the four GBT observations used to create the expected profile for Source 2 contain OH features centered near -40 km/s; the other GBT observation, at $l = 138.792^{\circ}, b = 2.0^{\circ}$ contains a double OH emission profile, neither peak of which is centered at -40 km/s (they are near -46 km/s and -38 km/s), so this spectrum was not used in the expected profile. Ignoring the uncertainty owing to unknown OH variation over the area, we obtain the results given in Table \ref{table:source2comps}.
\begin{table*}[ht!]
\begin{center}
\caption{Excitation temperatures for the components of Source 2. Errors from profile integrals and optical depth measurements. \label{table:source2comps}}
\begin{tabular}{c c c c c c}
\hline
 l & b & $T_C$ (K) & $T^{65}_{ex}$ (K) & $T^{67}_{ex}$ (K)\\
\hline
 138.792$^{\circ}$ & 2.284$^{\circ}$ &  3.8 &  4.9 $\pm$ 0.4  & 4.7 $\pm$ 0.3 \\
 138.650$^{\circ}$ & 2.142$^{\circ}$ &  3.9 &  4.9 $\pm$ 0.4    & 4.9 $\pm$ 0.3 \\
 138.934$^{\circ}$ & 2.142$^{\circ}$ &  3.9 &  4.6 $\pm$ 0.3    & 4.4 $\pm$ 0.2 \\                                    
\end{tabular}
\end{center}
\end{table*}

We again use the uncertainty of 2 K, and also recognize that $T_C$ = 3.9 K is a lower limit on $T_{ex}$. Taking a weighted mean of the results, we find for Source 2 alone: \\

3.9 K $  < T^{65}_{ex} < 6.8 $K, and   3.9 K$ < T^{67}_{ex} < 6.6 $ K, \\

\noindent and the mean of the results from Sources 1 and 2 is: \\

4.0 K $ < T^{65}_{ex} <$ 6.4 K, and  4.0 K $< T^{67}_{ex} <$ 6.2 K. \\

The results are consistent with those from the continuum background method. However, note that the expected profile method is unable to reveal a difference between the excitation temperatures of the two main OH lines in our data.

\section{Departures from the 5:9 ratio} \label{sec:LTE?}

A consequence of the differences which we have found in the main OH line excitation temperatures is that the ratio of the 1665:1667 main line intensities will differ from the value of 5:9, often referred to as the ``LTE'' value for optically-thin lines. Of course, high optical depth in the two lines would result in line ratios approaching 1:1. In this section we examine the GBT survey results in W5 for evidence of of such differences. Looking at the ratios of corresponding 1665 MHz and 1667 MHz OH emission lines in the GBT survey, for cases in which both lines occur as emission, one wonders if the emission is consistent with optically thin lines and a Boltzmann distribution energy level partition function. That is an assumption in the derivation of the equations for calculating OH column densities, so it is important to know if any pairs of 1665 MHz and 1667 MHz lines deviate from this partition function. In order to test whether the emission lines at a given coordinate are consistent with a Boltzmann distribution partition function, we start with Equation \ref{eqn:NOH} \cite[see e.g.][]{ll96}. This equation relates the column density of OH to $T_{ex}$, $T_C$, and the line profile integral. We can write this equation for either the 1665 MHz or the 1667 MHz line, each with its own set of parameters, but the version for 1665 MHz has a different coefficient from the version at 1667 MHz, and they each have a different value for $T_{ex}$ as well. Since the analysis using either line should yield the same value for the column density $N(OH)$, we can write:
\begin{eqnarray}
\label{eqn:NOH}
N(OH) & = & C_{67}\frac{T^{67}_{ex}}{T^{67}_{ex} - T_C}\int{\phi_{67}(\nu)d\nu}  \\    & = & C_{65}\frac{T^{65}_{ex}}{T^{65}_{ex} - T_C}\int{\phi_{65}(\nu)d\nu} \nonumber
\end{eqnarray}
\noindent where $C_{65}$ and $C_{67}$ are the coefficients for the equations at 1665 MHz and 1667 MHz respectively, and $T^{65}_{ex}$ and $T^{67}_{ex}$, and $\phi_{65}$ and $\phi_{67}$ are the excitation temperatures and line profiles at 1665 MHz and 1667 MHz respectively. We know that $C_{65} = (9/5)C_{67} = 1.8C_{67}$, so that:
\begin{equation}
\frac{T^{67}_{ex}}{T^{67}_{ex} - T_C}\int{\phi_{67}(\nu)d\nu} = 1.8\frac{T^{65}_{ex}}{T^{65}_{ex} - T_C}\int{\phi_{65}(\nu)d\nu}.
\label{eqn:LTE?}
\end{equation}
\noindent This equation should be satisfied if the OH energy levels are populated according to a Boltzmann distribution partition function. We find that for the majority of the OH detections in the grid survey with emission at both main line frequencies, Equation \ref{eqn:LTE?} is satisfied, indicating that the difference in $T_{ex}$ for the two lines is generally sufficient to explain why the lines are outside of the 1:1 - 5:9 ratio without invoking the possible additional effects of anomalous excitation, e.g.\ by IR photons.

Our observed absorption profiles are also in the 5:9 ratio with the exception of two profiles, which have more complicated structure. At $l = 136.875^{\circ}$, $b = 1.125^{\circ}$, there are two absorption lines at 1665 MHz and three at 1667 MHz, and at $l = 137.0^{\circ}$, $b = 1.1256^{\circ}$, there is one absorption line at 1665 MHz and two at 1667 MHz. This curiosity can be explained by noting that one of these OH features could be located farther back along the line of sight, in a shielded pocket in the middle of the HII region. We already noted that this could be true of the two brown-circled coordinates on Figure 5. In the case of the absorption profiles, the explanation could be that the OH feature farther back along the line of sight has $T_C \approx T^{65}_{ex}$ so the line does not appear at 1665 MHz, whereas at 1667 MHz it is observed in absorption along with the other OH profiles. If that is assumed to be the case, the other absorption lines are in a ratio between 1:1 and 5:9, an anomaly that could be attributed to optical depth effects.

\section{Conclusions}

Two methods were used to determine the excitation temperatures of the OH 18-cm main lines in W5. The continuum background method provides the most precise determination yet of the main line excitation temperature difference $T^{65}_{ex} > T^{67}_{ex}$, a $7 \sigma$ result, consistent with past findings \cite[e.g.][]{c79} but with greater significance. Absolute determinations of $T_{ex}$ in W5 are subject to greater uncertainties dominated by the unknown geometry of the OH distribution in the region, but both methods provide consistent results within the uncertainties, and the continuum background method yields the more precise results. Best estimates of the values are $T^{65}_{ex} = 6.0 \pm 0.5$ K, $T^{67}_{ex} = 5.1 \pm 0.2$ K, and $T^{65}_{ex} - T^{67}_{ex} = 0.9 \pm 0.5$ K., although if we avoid assumptions about simple geometry, these estimates are really upper limits with a minimum $T_{ex}$ value greater than 4 K. These $T_{ex}$ values are similar to those found in less active environments such as dark clouds by e.g.\ \cite{rwg76} or \cite{ll96}, and are also comparable to the average value of the distribution of $T_{ex}$ measurements reported by \cite{li18}. For a majority of the observed positions in the grid survey of W5, LTE line ratios between 1665 MHz and 1667 MHz hold if the differences in excitation temperatures for these two lines are first taken into account. The definitive confirmation of the result that $T^{65}_{ex} > T^{67}_{ex}$ using the continuum background method demonstrates the need for renewed consideration of possible physical explanations for this phenomenon, although providing such an explanation is beyond the scope of this paper.

\acknowledgments
\section{Acknowledgements}

We thank W. Miller Goss for his advice and assistance in this project, including his expertise on the subject of OH 18-cm observations and especially regarding the strategy for VLA observing, which was partly inspired by the observing strategy used by his former graduate student Claire Murray for a project involving a large survey for HI absorption using the VLA \cite[21-SPONGE,][]{m15}. We also thank Dr.\ Goss for discussions and suggestions about this paper, which led to improvements in its clarity. We thank the anonymous referee for many suggestions that have improved this paper. We are grateful to David E.\ Hogg, a collaborator and co-author on much of our research program, for his contributions to the overall project as well as many discussions, especially regarding the details of GBT observing and data reduction. The National Radio Astronomy Observatory is a facility of the National Science Foundation operated under cooperative agreement by Associated Universities, Inc. The Green Bank Observatory is a facility of the National Science Foundation operated under cooperative agreement by Associated Universities, Inc. The work reported here has been partially supported by the Director's Research Funds at the Space Telescope Science Institute. Support for this work was also provided by the NSF through the Grote Reber Fellowship Program administered by Associated Universities, Inc./National Radio Astronomy Observatory. The research presented in this paper has used data from the Canadian Galactic Plane Survey, a Canadian project with international partners, supported by the Natural Sciences and Engineering Research Council of Canada. This research has benefited from the SAOImage software provided by the Harvard-Smithsonian Center for Astrophysics; we are grateful to Bill Joye for his advice on smoothing the FCRAO CO data using SAOImage. 

\appendix

\section{Previous Measurements of OH 18-cm Excitation Temperatures} \label{app:Tex}

Several measured values for the OH main line excitation temperatures are collected here from the literature. These values have all been determined using the ``expected profile'' method.
\begin{table*}[h!]
\begin{center}
\caption{Values of OH main line excitation temperatures reported in selected past publications.}
\begin{tabular}{l l l} \hline
\hline
Publication & $T^{65}_{ex}$ (K) & $T^{67}_{ex}$ (K) \\
\hline
Heiles (1969)  & 4.5 to 9.6 & 4.5 to 9.6 \\ 
Manchester and Gordon (1971) & $T_C + 0.0 \pm 0.17$&$T_C + 0.45 \pm 0.14$ \\
Turner (1973) & likely higher in 6 cases & 3.71 to 12.03 \\
Heiles and Gordon (1975) & & $T_C$ + (0.13 $\pm 0.02$ to 1.40 $\pm 0.81$) \\
Nguyen-Q-Rieu et al. (1976) & $\approx$ 6.5 & $\approx$ 5.0 \\
Crutcher (1977) & $6.0 \pm 0.3$ to $7.0 \pm 0.5$ & $5.0 \pm 0.3$ \\
Crutcher (1979) & $4.2 \pm 0.3$ to $ > 6.7 \pm 0.4$ & $3.5 \pm 0.2$ to $ > 8.5 \pm 0.4$ \\
Liszt and Lucas (1996) & & $T_{CMB} + (0.2 $ to $2.2)$   \\
Li et al. (2018) & & distribution$(T_{ex}) \propto \frac{1}{\sqrt[]{2\pi}\sigma} \exp{\big(-\frac{ln(T_{ex}) - ln(3.4 K)^2}{2 \sigma^2}} \big)$ \\
\end{tabular}
\end{center}
\end{table*}
%

%






\end{document}